\documentclass[prb,twocolumn,superscriptaddress,showpacs,preprintnumbers,amsmath,amssymb,floatfix]{revtex4-1}

\usepackage{graphicx,tabularx,multirow}
\usepackage{dcolumn}
\usepackage{bm}
\usepackage{upgreek}
\usepackage{color,soul}

\newcommand{\eq}[1]{Eq.~(\ref{eq:#1})}

\newcommand{\figurenames}{Figs.~}

\begin{document}

\title{Local Magnetic Measurements of Trapped Flux Through a Permanent Current Path in Graphite}

\author{Markus Stiller} \email{markus@mstiller.org} 

\author{Pablo D. Esquinazi}

\author{Jose Barzola-Quiquia}

\author{Christian E. Precker} \affiliation{Division of
  Superconductivity and Magnetism, Felix-Bloch Institute for
  Solid-state Physics, University of Leipzig, 04103 Leipzig, Germany}

\date{\today}

\begin{abstract}
  Temperature and field dependent measurements of the electrical
  resistance of different natural graphite samples, suggest the
  existence of superconductivity at room temperature in some regions
  of the samples.  To verify whether dissipationless electrical
  currents are responsible for the trapped magnetic flux inferred from
  electrical resistance measurements, we localized them using magnetic
  force microscopy on a natural graphite sample in remanent state
  after applying a magnetic field.  The obtained evidence indicates
  that at room temperature a permanent current flows at the border of
  the trapped flux region. The current path vanishes at the same
  transition temperature $T_c\approx370$~K as the one obtained from
  electrical resistance measurements on the same sample.  This sudden
  decrease of the phase is different from what is expected for a
  ferromagnetic material. Time dependent measurements of the signal
  show the typical behavior of flux creep of a permanent current
  flowing in a superconductor.  The overall results support the
  existence of room-temperature superconductivity at certain regions
  in the graphite structure and indicate that magnetic force
  microscopy is suitable to localize them. Magnetic coupling is
  excluded as origin of the observed phase signal.
\end{abstract}

\maketitle

\section{Introduction}
\label{intro}

The usual way to probe the existence of superconductivity in a
material is by measurement of \textit{zero} resistance and magnetic
flux expulsion below a critical temperature. Experimentally, it is a
challenge to prove the existence of superconductivity in very small or
in granular regions of a macroscopic sample because typical experimental
methods, trying to show nominally \textit{zero} electrical resistance
and/or magnetic flux expulsion, are not well suitable. This is the case
where granular superconducting regions are localized within embedded
two-dimensional (2D) interfaces, at which no easy access for direct
electrical contacts to the regions of interest is possible. Moreover,
if the size of the superconducting regions is much smaller than the
effective London penetration depth $\lambda_L$, in addition to the
large demagnetization effects expected for 2D interfaces, the flux
expulsion, i.e.~the Meissner effect, is, strictly speaking,
negligible. In any case, a true zero resistance cannot be measured
using standard electric current/voltage measurements simply because
this would imply using devices with infinite sensitivity.

An alternative proof for the existence of superconductivity can rely
on the observation of dissipationless currents that maintain a
magnetic flux trapped at certain regions of a sample.  Early works
using magneto-optical flux imaging (MOI), determined the spatial
distribution and the magnitude of the supercurrents in high
temperature superconducting YBCO crystals.  Recently published
results~\cite{PECQZL16} suggest that some regions in natural graphite
samples show a superconducting-like transition at unexpected high
transition temperatures of $T_c\approx370$~K. The observation of Bragg
peaks in X-ray diffraction (XRD) measurements, which correspond to two
possible stable stacking orders, i.e.~rhombohedral and Bernal
graphite, suggests, that the interfaces are the regions where
high-temperature superconductivity can be
localized~\cite{MCP13,V13,PT15,HV16} due to the existence of flat
bands, which was predicted in theoretical
works~\cite{FLMNNH12,CFPEZ13,PSHGBC15}. 

The record temperature for superconductivity at $203$~K reported
recently in a sulfur hydride system at high
pressures~\cite{drozdov_2015} seems to be consistent with the
Bardeen-Cooper-Schrieffer (BCS) theory. A Van Hove singularity was
suggested as a possible reason for high-temperature
superconductivity~\cite{bianconi_2015}.  The resulting
Khodel-Shaginyan flat bands~\cite{yudin_2014} with dispersionless
energy relation~\cite{khodel_1990,volovik_fermi_1994} may lead to
superconductivity at high temperatures. Such flat bands also exist at
the surface of rhombohedral
graphite~\cite{kopnin_high-temperature_2013,pierucci_evidence_2015,hyart_two_2018}
or at the interfaces between rhombohedral and Bernal
graphite~\cite{munoz_tight-binding_2013}. The low critical temperature
for usual superconductors is a result of the exponential suppression
in the BCS equation for quadratic dispersion relations. In the case of
flat bands, a critical temperature orders of magnitude larger can be
expected, assuming similar Cooper pair interaction strengths. In such
a case, the critical temperature is proportional to the pairing
interaction strength and to the area of the flat band in momentum
space. Thus, at certain interfaces between rhombohedral and Bernal
graphite~\cite{HV16,kopnin_high-temperature_2013,munoz_tight-binding_2013},
twisted Bernal
layers~\cite{esquinazi_superconductivity_2014,san-jose_helical_2013},
or regions under strain~\cite{kauppila_flat-band_2016},
superconductivity with a critical temperature at or above room
temperature might be triggered.  In the last 43 years, hints for the
existence of superconductivity at very high temperatures in
graphite-based samples were
reported~\cite{antonowich_effect_1975,antonowicz_possible_1974,PECQZL16,EPSCQS17,kawashima_possible_2013,kopelevich_ferromagnetic_2000}.

In this study we used local magnetic force measurements to find the
region of trapped magnetic flux, due to a permanent current, in a
natural graphite sample. The temperature dependence and time decay of
the phase signal was monitored. The obtained results rule out magnetic
order as origin and support the existence of superconductivity at room
temperature at certain regions of graphite.

\section{Superconductivity and Zero Resistance}
\label{sec:sup1}

Superconductors exhibit characteristic properties, such as ideal
diamagnetism, quantized magnetic flux lines or the vanishing of the
electrical resistance. In this study, the main attention will be
paid to the latter feature, because it is almost impossible to
measure the first two effects on bulk samples where superconductivity
is confined to \textit{some} interfaces embedded in the bulk
material. On one hand, the quasi-two dimensional (2D)
superconductivity at interfaces, and due to the huge demagnetization
factor, impedes to a very large extent the expulsion of a field applied
normal to the superconducting area. On the other side, the effective
penetration depth $\Lambda = 2\lambda_L^2 /d_i$~\cite{pea64}, where
$\lambda_L$ is the London penetration depth and $d_i$ the thickness of
the superconducting interface, gets easily larger than the sample
size, making a direct imaging of a single vortex impossible.

\paragraph{Zero electric resistance}

Since superconductivity was observed first in mercury, the magnitude
of the decrease of the resistance, when crossing over to the
superconducting state, was an important question. At that time,
standard methods were used to measure the electrical resistance,
i.e.~the voltage drop was monitored across a current carrying
wire. Hence, it was only possible to state that the resistance dropped
below the sensitivity limit of the measurement device, which makes it
in principle impossible to prove, even nowadays, that the resistance
vanishes and it is exactly equal to zero. In 1911, Kammerlingh-Onnes
reported that ``while the resistance at $13.9$~K is still 0.034 times
the resistance of solid mercury extrapolated to $0^\circ$C, at $4.3$~K
it is only 0.00225, while at $3$~K it falls to less than
$0.0001$''~\cite{D84}. New experiments later that year showed that
between $4.21$~K and $4.19$~K the resistance dropped from
$0.115~\Upomega$ to less than $10^{-5}~\Upomega$.

Dealing with the problem of measuring very small resistances, already
in 1914 Kammerlingh-Onnes used a technique which is superior to
standard resistance measurements. He measured the decay time of an
induced current in a closed superconducting loop made of lead. For
this purpose, at first the ring is held in the normal state,
i.e.~above the transition temperature $T_c$. A permanent magnet was
used to apply a magnetic field, inside the ring. The ring was then
cooled below $T_c$ to $1.8$~K, the magnetic field inside the
superconducting ring remains unchanged. The magnet was removed, thus
inducing a current. Kammerlingh-Onnes used a compass needle placed
close to the superconducting ring in order to measure any changes in
the magnetic field and thus, in the current flowing through the
superconducting ring. He then reported that within an hour, the
current (0.6~A) did not decrease, indicating that the resistance has
zero value and that the current would continue to flow as permanent
current.

In this way, it is possible to estimate a new upper limit for the
resistance, which is much more accurate compared to the electrical
resistance measurements~\cite{KB16}.  With $U_{\rm i}$ being the
induced voltage, the self-induction $L$ can be defined as
$U_{\rm i}=-L(\mathrm{d}I/\mathrm{d}t)$, and the stored energy of a
ring with permanent current is $(1/2)LI^2$
($\mathrm{d}P=LI\mathrm{d}I$). The change of this energy with time is
equal to the heating power $RI^2$, thus
\begin{equation}
  \label{eq:Ttdeo2}
  -LI\frac{\mathrm{d}I}{\mathrm{d}t}=RI^2.
\end{equation}
Hence $-(\mathrm{d}I/\mathrm{d_w}t)=(R/L)I$ with solution:
\begin{equation}
  \label{eq:Itdep}
  I(t)=I_0\exp\left(-\frac{Rt}{L}\right),
\end{equation}
where $I_0$ is the initial current at time $t=0$. This implies that
the decay of the current depends on the shape of the superconducting
loop. For a circular ring, the self-induction is given
as~\cite{Elliot99} $L=\upmu_0r\left[\ln(8r/d)-1.75\right]$. Assuming
that the radius of the ring is $r=300~\upmu$m, the radius of the wire
$d_w=0.5~\upmu$m, and that the current decreases less than 1~\% per
hour, then the resistance must be smaller than
\begin{equation}
  \label{eq:Restimate}
  R\leq\frac{-\ln(0.99)\cdot2.53\mathrm{nH}}{3600\mathrm{s}}\approx7.1\times10^{-15}~\Upomega.
\end{equation}
This shows that, monitoring the magnetic fields created by a permanent
current is much more accurate to estimate an upper limit of the
resistance, compared to standard measurements of the electrical
resistance. In the first experiments a compass needle was used, later
more sensitive methods were employed, such as torsion thread
experiments or magnetic force microscopy. However, in some cases the
resistance does not vanish, e.g.~small currents if magnetic flux
lines exist or alternating currents.

\section{Magnetic Force Microscopy}
\label{sec:mfm}
\begin{figure}
  \includegraphics[width=\columnwidth]{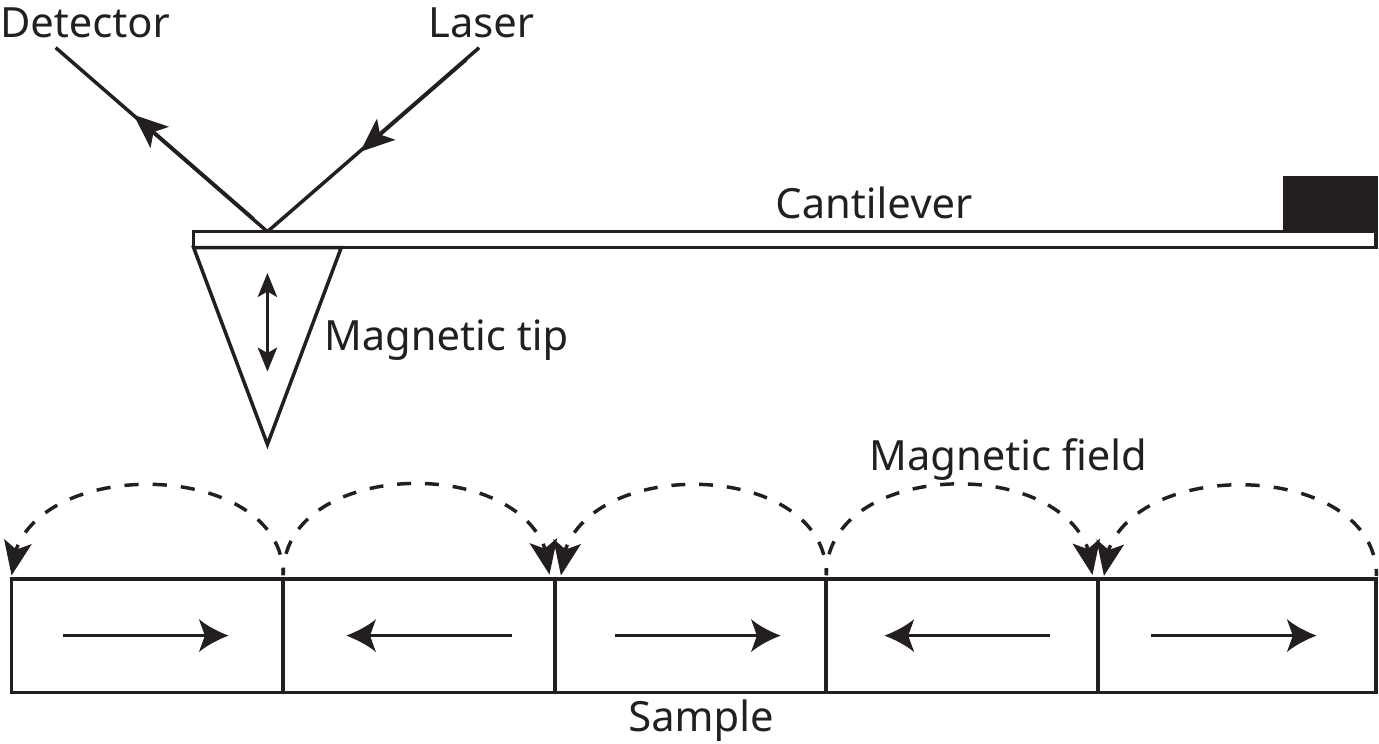}
  \caption{Basic principle of magnetic force microscopy setup. The
    magnetic tip is used to detect the magnetic stray fields of a
    sample. The sketch at the bottom depicts a sample with
    antiparallel ordered magnetic domains with the main magnetization
    parallel to the main area of the sample.}
\label{fig:mfm_basic}       
\end{figure}

\paragraph{Basic Principles}
In \figurename~\ref{fig:mfm_basic} the basic principle of magnetic
force microscopy (MFM) is shown. 
The sample, consisting of magnetic domains, emanates magnetic stray
fields, which are detected by a magnetized tip. In conventional MFM
devices, a laser is used to detect the oscillations of the cantilever,
however, self-sensing and self-actuating cantilevers are also
used~\cite{SQESTMA17}. Assuming the tip as a point dipole, then the
force acting on the tip is given by
\begin{equation}
  \label{eq:mfm1}
  \vec F=\upmu_0\nabla\left(\vec m\cdot\vec H\right),
\end{equation}
where $\upmu_0$ is the magnetic permeability of free space, $\vec m$
is the magnetic moment of the tip and $\vec H$ the magnetic stray
field from the sample at the position of the tip. 

\paragraph{Measurement technique} In MFM, non-contact mode is used to
detect the stray fields, i.e.~the distance between tip and sample
surface is kept at a constant value. To do so, for each line scan, the
topography is measured using tapping mode prior to the actual
non-contact measurement. In tapping mode, the cantilever is
oscillating, and the tip is deflected when encountering the sample
surface. A line scan is performed and thus the topography of the
sample is measured. In a second, non-contact scan, the same line is
followed but with the tip kept at constant distance $d$.

Assuming that the tip is a point dipole, the cantilever is
parallel to the sample surface and that the tip and sample are
independent of each other, then the force and force derivative
are~\cite{WG92}
\begin{align}
  \label{eq:mfm2}
  F &= m_x\frac{\partial B_x}{\partial z}+m_y\frac{\partial B_y}{\partial z}+m_z\frac{\partial B_z}{\partial z}, \\ 
  F'&= m_x\frac{\partial^2 B_x}{\partial z^2}+m_y\frac{\partial^2 B_y}{\partial z^2}+m_z\frac{\partial^2 B_z}{\partial z^2},
\end{align}
where $m_i$, $i=x,y,z$, are the effective magnetic moments of
tip. Assuming further an infinitely long dipole tip, and if the tip
magnetization is perfectly aligned along $z$, then~\cite{WG92}
\begin{align}
  \label{eq:mfm3}
  F &= m_x\frac{\partial B_z}{\partial z}, \\ 
\label{eq:mfm4}
  \varphi=-\frac{Q}{k}\left(\frac{\partial F}{\partial z}\right)&= -\frac{Q}{k}\left(m_z\frac{\partial^2 B_z}{\partial z^2}\right),
\end{align}
where $Q$ and $k$ are the quality factor of the tip resonance peak
and spring constant, respectively.

\paragraph{Experimental Details}

A commercial tip ($k\simeq3$~N/m) with a Cobalt-Chromium coating and a
nominal tip radius of 35~nm was used, which was placed in the MFM
device (Nanoscope). The graphite sample was fixed on a substrate (Si
with $150$~nm SiN$_x$ top layer) using varnish. The contacts (four
terminal sensing) for the electrical resistance measurement were done
using silver paste and gold wires. The resistance measurements were
carried out in a magneto-cryostat with a temperature stabilization of
a few mK at 300~K, for more details see~\cite{PECQZL16}. After the
resistance measurements, the sample was heated up to 390~K, so that
the trapped magnetic flux vanishes, followed by a zero-field cool to
room temperature. Substrate, including sample, was then
placed and fixed with varnish on a copper plate, where a thermometer
and heater were at the backside of the plate. The sample and copper
plate were connected to ground, to avoid electrostatic influences. MFM
measurements were then carried out in the usual way.

\begin{figure}
  \includegraphics[width=\columnwidth]{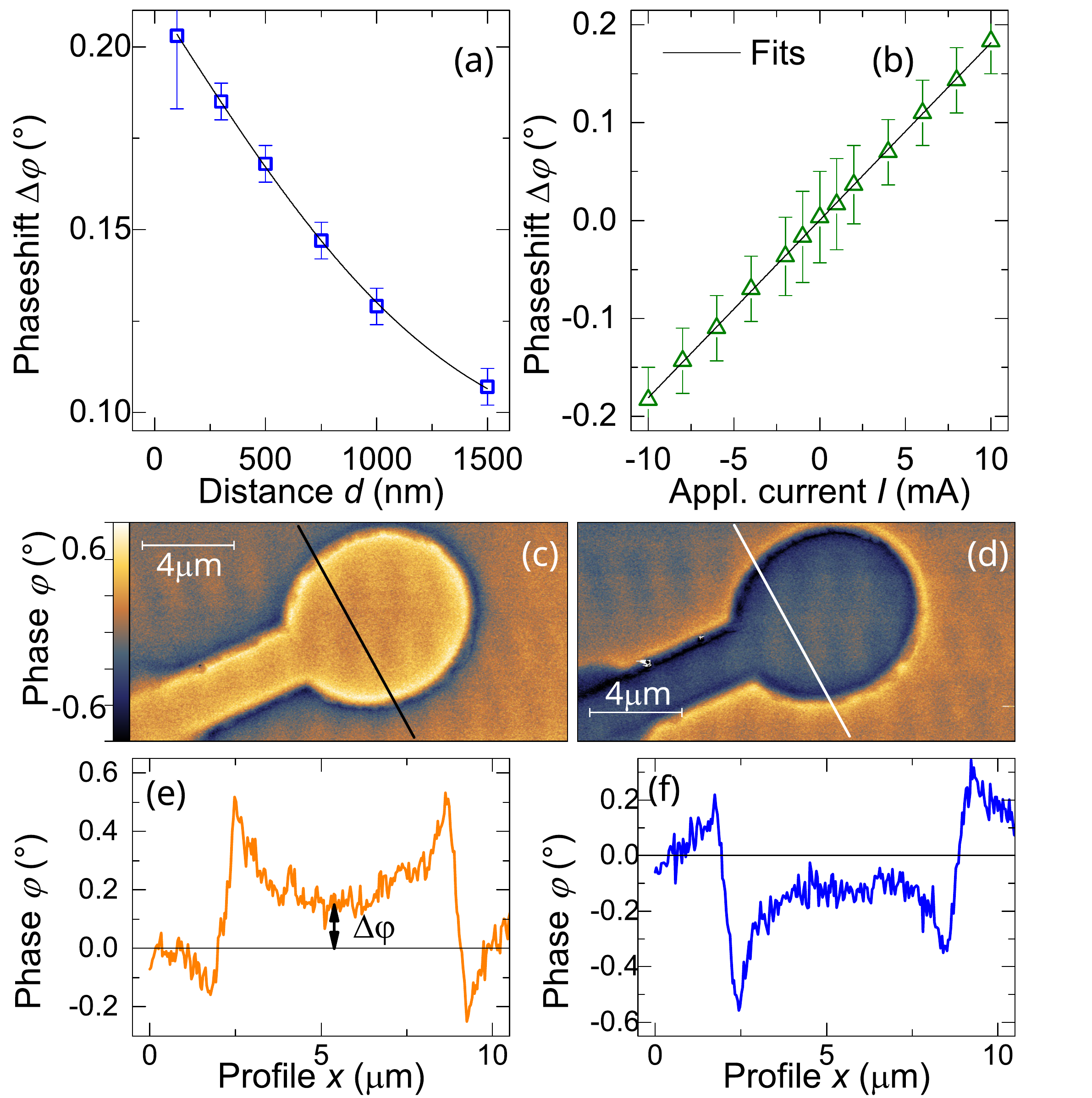}
  \caption{The phase shift $\Updelta\varphi$ for a current loop,
    depending on (a) the distance $d$ and (b), the applied current
    $I$. In (a), 10~mA was applied and (b) was measured at a distance
    of 200~nm. The lines are the fits to \eq{mfm6}, (c) and (d) are MFM
    images with $I=\pm3$~mA, (e) and (f) are the corresponding
    spectra.}
\label{fig:tipcali}       
\end{figure}

\paragraph{Characterization of the tip using a current loop}
According to \eq{mfm4}, we need to calculate the second derivate of
the field produced by a current loop. For simplicity, the following
substitutions are used~\cite{SLIY01}:$\rho^2=x^2+y^2$,
$r^2=x^2+y^2+z^2$, $\alpha^2=a^2+r^2-2a\rho$,
$\beta^2=a^2+r^2+2a\rho$, $k^2=1-\alpha^2/\beta^2$, $C=\upmu_0I/\pi$,
where $I$ is the current through the loop, $a$ is the radius and
origin is placed at the center of the loop, in the $x-y$ plane. The
$z$-component of the magnetic field is then given by~\cite{SLIY01}
(assuming the cross section of the conducting path is negligible)
\begin{equation}
  \label{eq:mfm5}
  B_z=\frac{C}{2\alpha^2\beta}\left[\left(a^2-r^2\right)E(k^2)+\alpha^2K(k^2)\right],
\end{equation}
where $K$ and $E$ are elliptic integrals of first and second kind,
respectively. The phase can now be simulated using
Eqs.~(\ref{eq:mfm4}) and~(\ref{eq:mfm5}), the calculations were
carried out using \textit{Mathematica}$^{\rm TM}$.  Further, within
the dipole approximation, the phase at the center of the loop is then
\begin{equation}
  \label{eq:mfm6}
  \Updelta\varphi=\frac{3\upmu_0a^2Im_zQ}{2k}\left[\frac{a^2-4(d+\delta)^2}{(a^2+(d+\delta)^2)^{7/2}}\right],
\end{equation}
where $d$ is the lift scan height, $\delta$ is the tip-dipole distance
and $\Updelta\varphi$ is the difference between the value at the
center of the loop and the value at the edge of the measured spectra,
i.e.~where $\varphi$ is approximately constant, see
\figurename~\ref{fig:tipcali}(e). Note that the shown line scans are
taken from the images, the scans used to calibrate the tip were
measured across the center of the loop such that there was sufficient
space for the phase to saturate. Further, seven lines were saved and
used to obtain an averaged scan.
We measured $\Updelta\varphi$ as function of distance $d$ and applied
current $I$, the data and the fits are shown in
\figurenames~\ref{fig:tipcali}(a) and (b). The fits yield
$m_z=(1.27\pm0.2)\times10^{-13}$~Am$^2$ and
$\delta=(1.31\pm0.2)~\upmu$m, these values agree with what has been
observed in the literature~\cite{KC97,LKCDW99,SQESTMA17}. However, it is
important to be aware that these are effective results and might
differ for different samples. This is due to the different
magnetic decay lengths of the magnetic field, which results in a
different effective magnetic volume of the tip within the field of the
sample. MFM images of another current loop with applied currents of
$\pm3$~mA are shown in \figurenames~\ref{fig:tipcali}(c) and (d), the
lines indicate the position of the phase spectra shown in
\figurenames~\ref{fig:tipcali}(e) and (f).

\paragraph{MFM of ferromagnetic samples}

MFM is a powerful tool to characterize the magnetic stray fields of a
variety of samples, such as thin films or nano/micro-structures. This
is especially interesting for cases where other measurements
techniques are not suitable, e.g.~a SQUID is not practical to measure
samples with a large background due to substrates or, as it is the case
for interfaces in graphite, due to the bulk contribution.
\begin{figure*}
  \includegraphics[width=.75\textwidth]{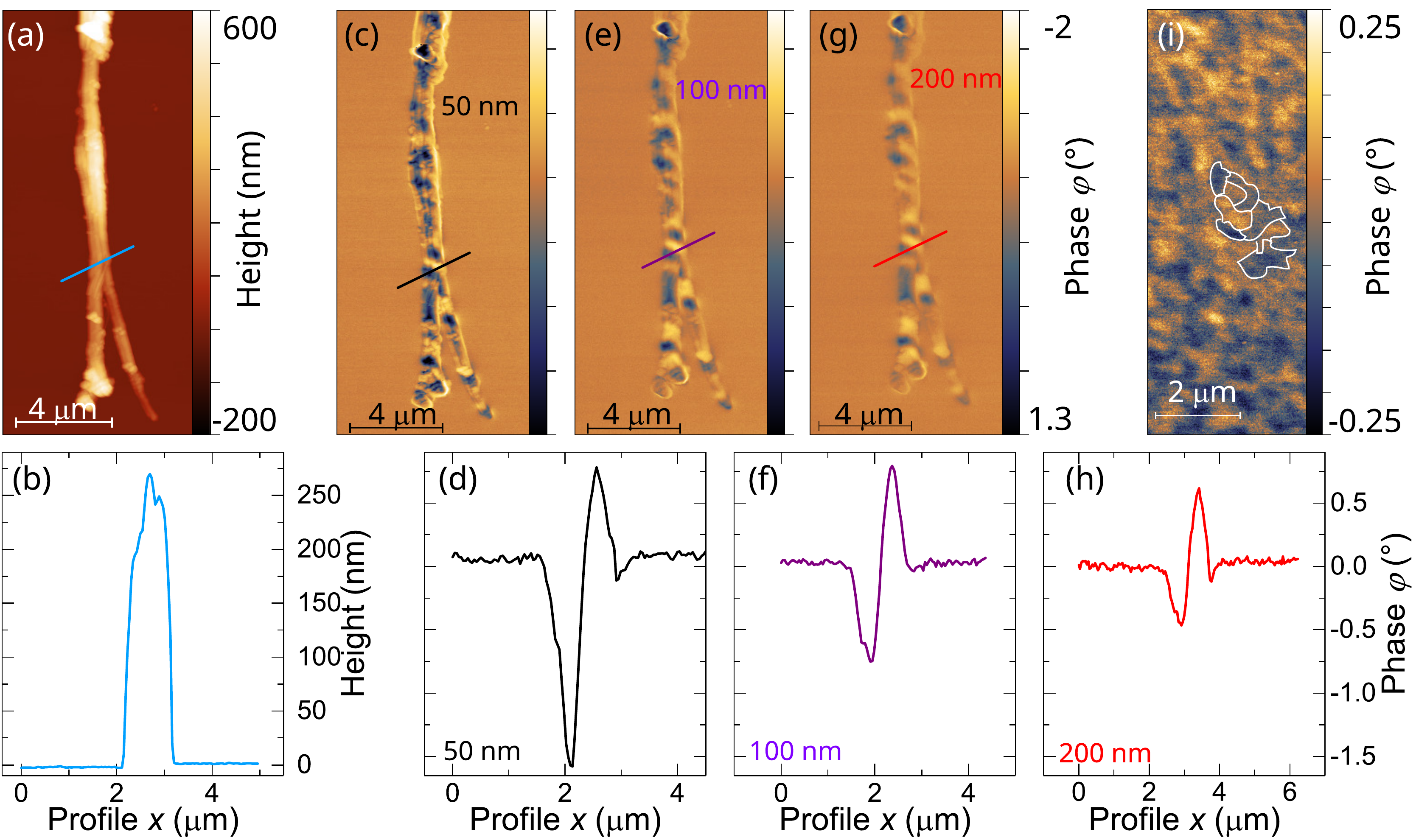}
  \caption{The topography ((a)\&(b)) and the measured phase ((c)-(h))
    of a multi-walled carbon nanotube filled with Fe at different lift
    heights. (i) is the MFM image of a ZnFe$_2$O$_4$ ferromagnetic
    thin film.}
\label{fig:mwcnt}       
\end{figure*}
As an example of typical MFM images of ferromagnetic samples, the
measurements of a multi-walled carbon nanotubes (MWCNT) filled with
iron, see~\figurenames~\ref{fig:mwcnt}((a)--(h)), and a ZnFe$_2$O$_4$
ferromagnetic thin film,~\figurename~\ref{fig:mwcnt}(i), are
shown. Both samples show clear magnetic domains and domain walls
between them. Both samples were not magnetized prior to the MFM scans,
thus the domains are pointing in arbitrary
directions. \figurename~\ref{fig:mwcnt}(a) shows an image of the
topography of the nanotube, \figurename~\ref{fig:mwcnt}(b) the
corresponding line scan. The MWCNT was measured at three different
scan heights, 50~nm, 100~nm and 150~nm, the images
(\figurenames~\ref{fig:mwcnt}(c),(e),(g)) and line scans
(\figurenames~\ref{fig:mwcnt}(d),(f),(h)) are given. The phase line
scan profiles show N{\'e}el wall features and beside the MWCNT the
phase is zero, where no magnetic field is present (or it is a
constant).

\section{Results}
\label{sec:results}

In this Section we will present the results of measurements on the
graphite sample. In Section~\ref{sec:edx} the result of
energy-dispersive X-ray spectroscopy is shown. A comparison of the
sample before and after applying a magnetic field is given in
Section~\ref{sec:befaft}, an investigation on the temperature
dependence on the phase signal can be found in
Section~\ref{sec:TempDep}. Further, the change over an extended period
of time is presented in Section~\ref{sec:fluxcreep}.

\subsection{Energy-dispersive X-ray spectroscopy}
\label{sec:edx}

\begin{figure}
  \includegraphics[width=\columnwidth]{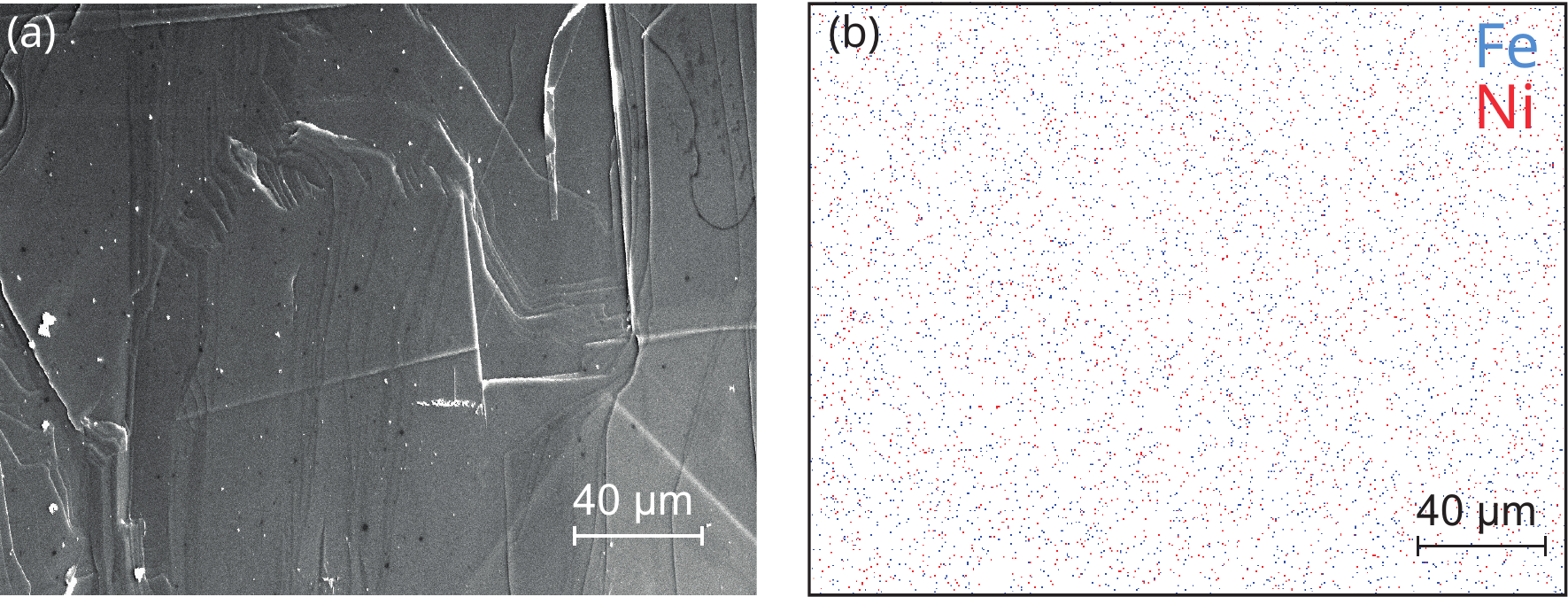}
  \caption{(a) Scanning electron microscopy image of the region where
    the current line is located. (b) is the corresponding
    energy-dispersive X-ray image for iron and nickel.}
\label{fig:edx}       
\end{figure}

In order to check for the presence of rhombohedral and Bernal
graphite, X-ray diffraction measurements were performed at the initial
material, the results can be found in~\cite{PECQZL16}. The existence
of both, Bernal and rhombohedral graphite was confirmed. To check for
the presence of magnetic impurities, energy-dispersive X-ray (EDX)
spectroscopy was carried out at the position of the sample, where the
permanent current path was measured. The results of a map scan are
shown in \figurename~\ref{fig:edx}.
The EDX measurement yields a carbon content of $\sim$94 At~\% and
$\sim$6 At~\% of oxygen. \figurenames~\ref{fig:edx}(a) and (b) show
the SEM image and the results for Fe and Ni, respectively. We find
isolated, homogeneously dispersed impurities, with less than
0.004~At~\% according to the EDX software analysis. There is no
topographic structure resembling the phase signal and the amount of
impurities is too low for ferromagnetic coupling. Therefore, we can
rule out magnetic impurities as origin of the phase signal.

\subsection{Before/After application of an external magnetic field}
\label{sec:befaft}

Before we started the MFM measurements, the sample has been put into
the virgin state. It means, that the sample was placed in the
magnetocryostat, for convenience and also to measure the electrical
resistance, see Section~\ref{sec:fluxcreep}. Then it was heated to
$T=390$~K followed by a cool down at no field to room
temperature. The so-prepared sample was measured for several weeks, in
order to cover a large area.
\begin{figure}
  \includegraphics[width=\columnwidth]{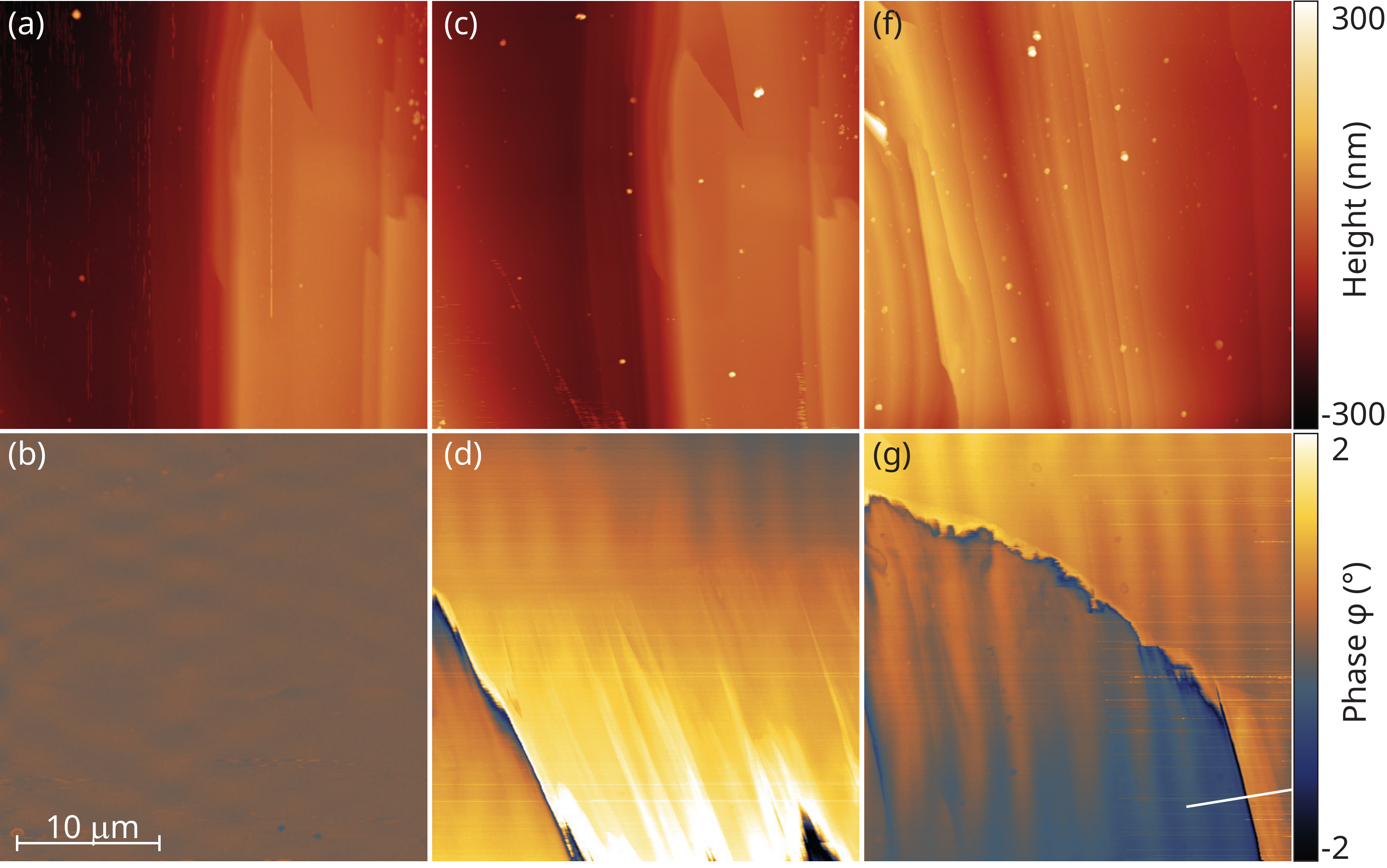}
  \caption{Topography images (top) and phase images (bottom) before
    ((a)\&(b)), and after ((c)\&(d)) application of an external
    magnetic field; (f) and (g) show the
    topography and phase at another position.}
\label{fig:befaft}       
\end{figure}
A MFM image is shown in \figurename~\ref{fig:befaft}(b), and, as it can
be seen, there is no sign of magnetic domains. If there was
ferro- or ferrimagnetic order with the corresponding magnetic domains,
all domains would have a spontaneous orientation and there would
be a change in the phase. If the domains were smaller than the lateral
resolution, i.e.~50~nm, we would see some unresolved averaged changes
in the phase, but changes nevertheless.

In the next step, a permanent magnet (for $\approx 10$~s, magnetic
field $\approx0.05$~T, measured with a Hall sensor) was placed near
the sample with the magnetic field oriented perpendicular to the
sample surface. After application of the external magnetic field, the
MFM measurements were continued. In \figurename~\ref{fig:befaft}(d)
the result is shown at the same position as in
\figurename~\ref{fig:befaft}(b), as the corresponding topography
images indicate (\figurenames~\ref{fig:befaft}(a)--(d)). A clear phase
feature appears, \figurename~\ref{fig:befaft}(g) shows the phase image
at another position, where the feature continued and is clearly
visible. From the topography and phase images, it is obvious that
there is no relation between surface and phase signal. However, as the
naked eye can give a misleading conclusion, neural network training
(Gwyddion) was used to find possible correlations. In this method, a
trained network can process data to find a signal from a model. As
model, the topography and as signal, the phase was used to train a
network, which was then applied to models (topography) in order to
calculate the signal (phase). It was not possible to reproduce any
phase image this way, confirming that there is no relation between
surface and phase.

\begin{figure}
  \includegraphics[width=\columnwidth]{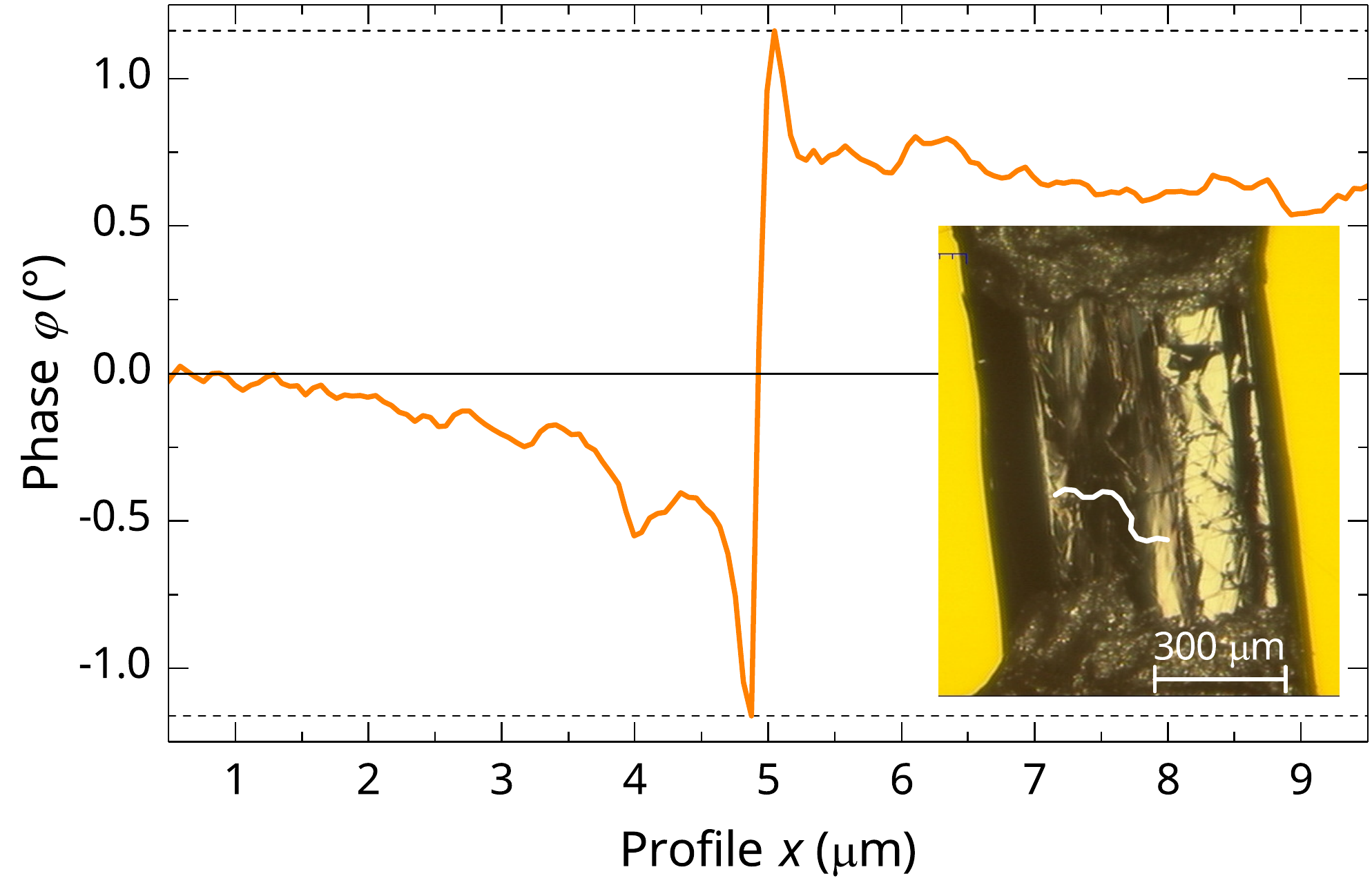}
  \caption{Line spectra indicated in \figurename~\ref{fig:befaft}(g),
    for comparison see the scan of the current loop in
    \figurename~\ref{fig:tipcali}(e). The optical image shows the
    sample and the position of the persistent current.}
\label{fig:spec}       
\end{figure}

A line scan, at the position indicated in
\figurename~\ref{fig:befaft}(g) as white line, is shown in
\figurename~\ref{fig:spec}; the inset is an optical image of the
sample, the line shows the position of the persistent current. We
could not measure the complete loop, only up to the edges of some
rough surface regions.
The line scan looks similar to that of a current loop, see
\figurename~\ref{fig:tipcali}(e) and compare with the scans of
magnetic domains, e.g.~\figurenames~\ref{fig:mwcnt}((d)--(h)). This
result strongly suggests that the observed signal is not due to
magnetic order: After the application of an external field, magnetic
domains would have been aligned in $z$-direction. It means that we
would need to have two areas, one magnetically ordered (with a single
domain) and one without any domain. But this would imply that magnetic
stray field existed from the beginning -- before applying an external
field -- because regardless of the magnetization direction and/or size
of the domain(s), at the boundary of the ferromagnetic/nonmagnetic
region, stray fields would have been present. As shown before, no
stray field signals were measured, neither at the shown position, nor
at any scan position of the sample in the virgin state.

\subsection{Temperature dependence}
\label{sec:TempDep}
\begin{figure*}
  \includegraphics[width=\textwidth]{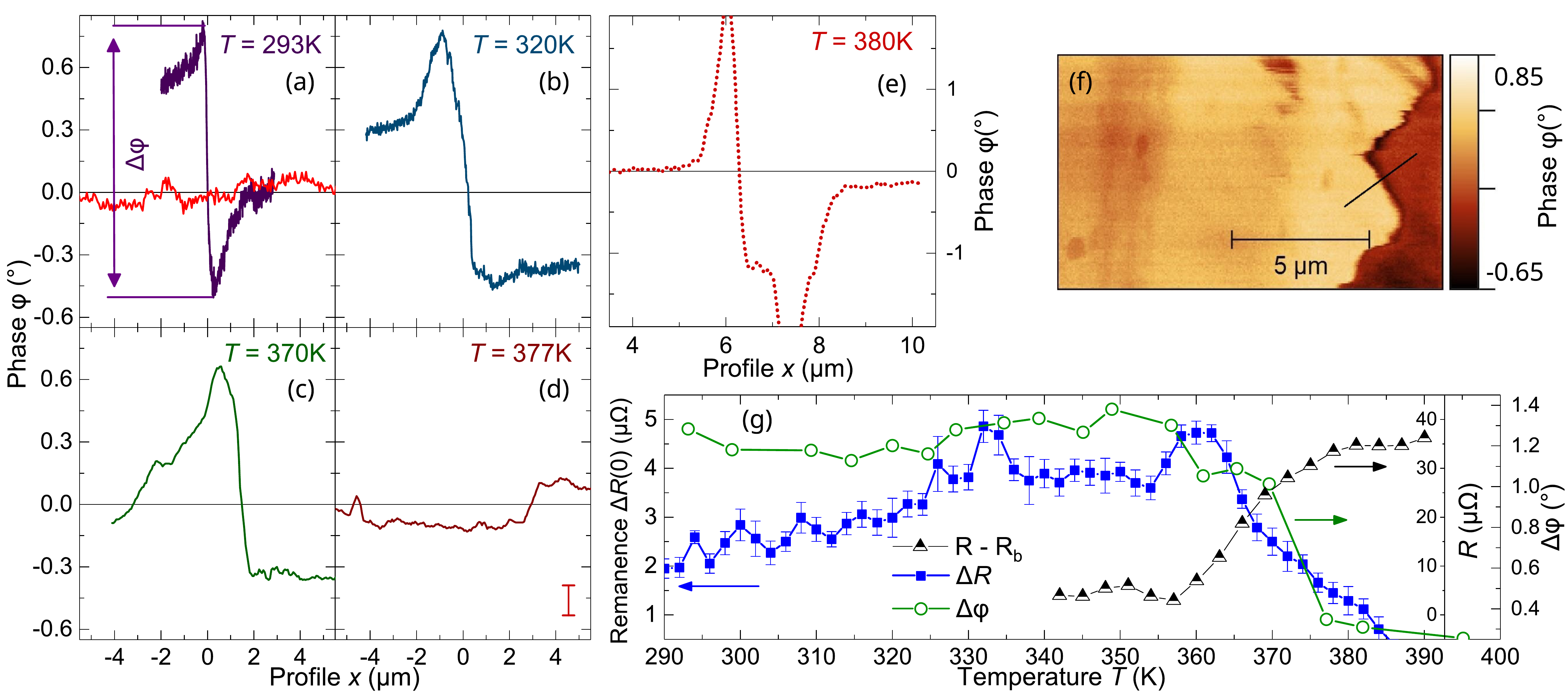}
  \caption{MFM line scans of the sample at different temperatures
    ((a)--(d)) at the position indicated in (f), (e) is a line scan
    using the current loop at $T=380$~K.  In (a), the purple line is
    the scan of the sample in remanent state (and used to define
    $\Updelta\varphi$), the red line is the result after heating the
    sample at $T=395$~K, and cool down back to room temperature. In
    (g), $\Updelta\varphi$, the resistance (after a linear background
    subtraction) and the remanence $\Updelta R(0)$~\cite{PECQZL16} as
    a function of temperature is given.}
\label{fig:tdepmfm}       
\end{figure*}

The temperature dependence of the resistance was already measured
before the MFM measurements and a transition was found at
$T\approx370$~K, in agreement with measurements done in similar
samples~\cite{PECQZL16}. For this reason, we measured the
temperature dependent MFM and the results are shown in
\figurename~\ref{fig:tdepmfm}.
In \figurenames~\ref{fig:tdepmfm}((a)--(d)) the lines scans at four
different temperatures can be seen. An optical images of the scanned
area is given in \figurename~\ref{fig:tdepmfm}(f). The scans show a
well defined offset between either side of the loop. This is due to
the shape of the loop at this position, which resembles a U-turn, with
a rather constant magnetic field in the interior. Several scans have
been measured (in temperature steps of 5~K), for convenience not all
are shown, however, they look very similar to the shown profiles up to
$T=370$~K. The magnitude of the phase remains constant, only a
broadening can be seen. A remarkable change occurs at $T=377$~K, where
a sudden decrease of the phase sets in, see
\figurename~\ref{fig:tdepmfm}(g). MFM measurements were continued to
$T=395$~K, and then the sample was zero-field cooled back to room
temperature. The phase signal did not reappear at any point, the
result at room temperature can be seen in
\figurename~\ref{fig:tdepmfm}(a) as red line.

The sharp transition around $T=377$~K contradicts the expected
behavior of ferromagnetism, where a continuous decrease of the
magnetic coupling would cause a continuous decrease of the magnitude
of the phase. Indeed, the decrease of the remanence with temperature
in graphite samples that show magnetic order, follows a nearly linear
decrease with temperature from low to about 0.9 of the Curie
temperature~\cite{QERSBG07}; at higher temperatures near $T_C$, the
decrease is even stronger.  This dependence is compatible with 2D spin
waves~\cite{QERSBG07} measured in non-irradiated~\cite{EQSRQG10} as
well as irradiated graphite~\cite{QERSBG07}, see also the review
in~\cite{chap3} for details and additional literature.

Further, when cooling without applied magnetic field, magnetic domains
would tend to order spontaneously in random directions, and a domain
structure similar as shown in \figurename~\ref{fig:mwcnt}(i) would
be expected. Yet, no signal in the phase was obtained after heating
above this critical temperature and cooling down again. Using
$\Updelta\varphi$, see \figurename~\ref{fig:tdepmfm}(a), we plot the
temperature dependence of the phase in
\figurename~\ref{fig:tdepmfm}(g). The transition is clearly visible
around $T=370$~K and the general behavior agrees very well with the
resistance measurements of the sample, i.e. $R(T)$ after a linear
background subtraction, and the remanence
$\Updelta R(0)=R_B(0)-R_0(0)$. Here, $R_B(0)$ is the resistance
measured at zero field after applying a field of 0.03~T normal to the
main sample surface, and $R_0(0)$ is the resistance of the sample in
the virgin state, obtained after zero-field cooling from $T=390$~K,
which was done for each temperature. For more details regarding the
temperature dependence of this remanence in different graphite
samples see~\cite{PECQZL16}.

The two resistance measurements and the MFM scans indicate a critical
temperature of $T_c\approx370$~K with a transition width of
$\lesssim20$~K. In \figurename~\ref{fig:tdepmfm}(e), the MFM line scan
of a current loop is shown, measured at $T=380$~K under the same
conditions as the measurements done for the graphite sample. This
confirms that the MFM experiments work at temperatures higher than
$T_c$, and that the tip remains magnetized at such elevated
temperatures. Note that, if the magnetization of the tip was
influenced by field of the current loop, we would have seen a
hysteresis-like shape in the phase scans (backward/forward scan
direction), which was not the case.

\subsection{Flux creep}
\label{sec:fluxcreep}

After applying an external magnetic field, it took approximately 22
days to find the position of the permanent current path and to
prepare to follow up with the experiments.
\begin{figure}
  \includegraphics[width=\columnwidth]{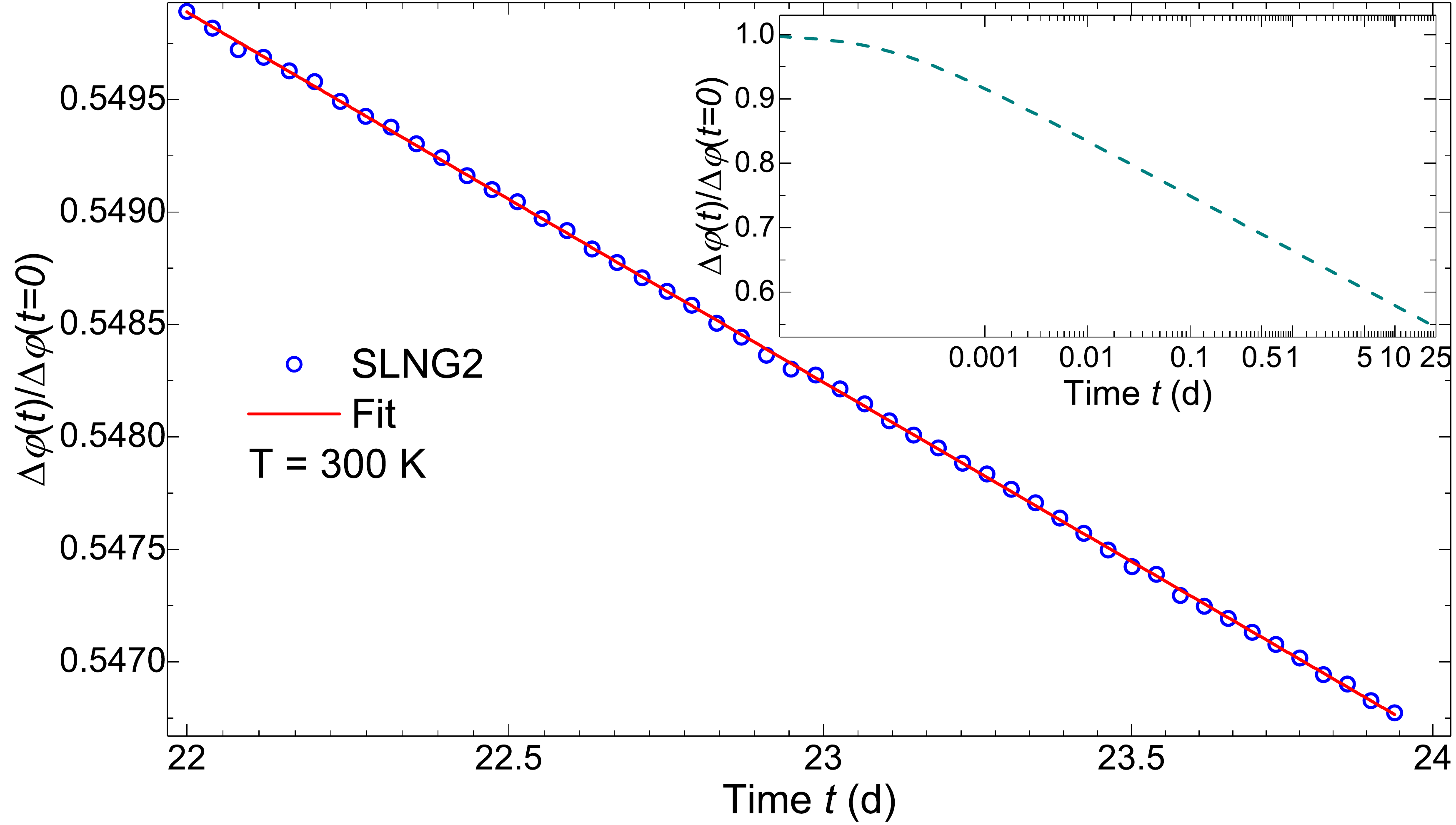}
  \caption{Normalized phase shift $\Updelta\varphi$ as a function of
    time, measured after a period of $\approx22$ days. The continuous
    line is the fit to \eq{fluxcreep}. The inset shows the resulting
    phase shift starting from $t\simeq0$ (dashed line).}
\label{fig:fluxcreep}       
\end{figure}
This period followed an investigation of the time dependence of the
phase shift $\Updelta\varphi$. For this purpose, a suitable spot was
chosen and measured for 2 days such that approximately every hour, a
$(10\times10)\upmu$m$^2$ image was obtained. The phase shift was then
obtained at the same position of the sample, considering the
piezo creep (of the positioners), especially within the first hours of
the experiment, see \figurename~\ref{fig:fluxcreep}. 

\paragraph{Estimate of the resistance} Using $\Updelta\varphi(T)$, an
effective electrical resistance can be estimated with \eq{Restimate}
that would produce the measured decrease in phase shift, i.e., in the
current amplitude.  Applying the parameters mentioned above, we find a
resistance of $\approx8\times10^{-17}$~Ohm. The intrinsic ohmic
resistance of the loop should be much smaller, eventually zero. The
current line remains for several weeks which indicates the existence
of a permanent current which originates the magnetic field and is also
the reason for the observed remanence in the resistance. However, the
reduction of $\Updelta\varphi$ can be explained in terms of
creep. Since
\begin{equation}
  \frac{\partial m}{\partial t}\propto\frac{\partial \phi}{\partial t}\propto\frac{\partial I}{\partial t}\propto\frac{\partial \varphi}{\partial t},
\end{equation}
the well known logarithmic time dependence~\cite{GK93} can be written
as
\begin{equation}
  \label{eq:fluxcreep}
\frac{\Updelta\varphi(t)}{\Updelta\varphi(0)}=1-\Updelta\varphi_1\ln\left(1+\frac{t}{\tau}\right),
\end{equation}
where $\tau$ is a time constant and $\varphi_1\propto k_{\rm B}T/U_a$
with $U_a$ being the flux creep activation energy. The results and the
fit to \eq{fluxcreep} can be seen in \figurename~\ref{fig:fluxcreep},
where $\tau=10$~s has been taken from~\cite{PECQZL16}. The inset shows
$\Updelta\varphi(t)$ starting from time $t\simeq0$, which was
calculated using the results of the fit ($\Updelta\varphi_1=0.87$).

The meandering-like structure of the measured current path is similar
to the one observed in high-temperature superconducting oxides, where
a modified Bean model, that includes the lower critical field
$H_{c1}$, was used to understand the origin of the Meissner
hole~\cite{VWCGKN97,VWCGKN98,JAKLK02}. However, if superconductivity
is localized at the interfaces of the graphite sample~\cite{PECQZL16},
from the Ginzburg-Laundau relation for $H_{c1}\propto1/\Lambda^2$ and
using $\Lambda = 2\lambda_L^2 /d_i$, one expects $H_{\rm c1}$ to be
negligible due to the huge penetration depth. Further, the pinning of
the pancake vortices within the interfaces would also be
negligible. In contrast, the here presented sample as well as similar
ones, show a maxima in the remanence of the resistance
$\Updelta R(0)$ not far from the critical temperature. This indicates
that the magnetic field at remanence is produced by macroscopic (or
mesoscopic) current loops, i.e.~fluxons, not pinned pancake vortices.
These fluxons are the origin for the remanent magnetic field and the
irreversible behavior observed in the electrical resistance and
magnetization.
\begin{figure*}
  \includegraphics[width=.7\textwidth]{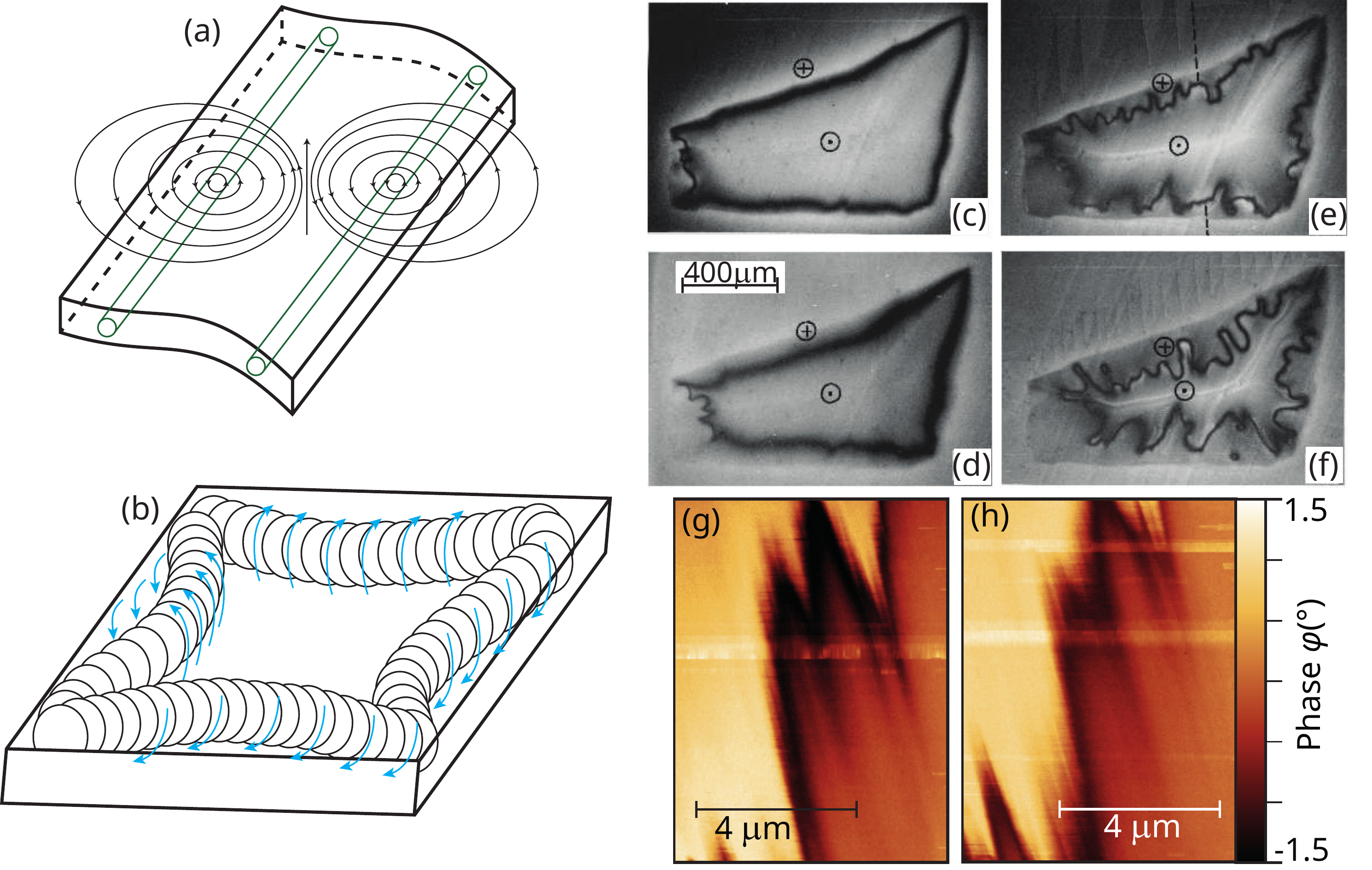}
  \caption{Figures ((a)--(f)) are adapted
    from~\cite{VWCGKN97,VWCGKN98}. (a) shows a sketch of magnetic
    field of a superconducting plate in remanent state. The Meissner
    holes with $B=0$ are shown, the shape (b) in a square-shaped
    plate, critical currents in the bulk and Meissner current along
    the boundary of the flux-free cylinder. Magnetic flux patterns in
    a 40$~\upmu$m thick YBCO crystal in: (c) remanent state after
    field cooling in $\upmu_0H=150$~mT and switching off the field at
    $T=20$~K; (d) application of $\upmu_0H=-100$~mT after (c), the
    remagnetization front moved further inside the sample; (e)
    remanent state after field cooling in $\upmu_0H=150$~mT and
    switching off at $T=55$~K; (f) application of $\upmu_0H=-23$~mT
    after (e). (g) and (h) are MFM images of the graphite samples with
    a time difference of 2 days, the drift is position was
    corrected. The time-dependent creep of fluxons can be seen. The
    shape of the fluxons resembles the shape of the Meissner holes in
    (e) and (f), where also the boundary between two flux directions
    is narrow and strongly bent.}
\label{fig:fluxcreep2}
\end{figure*}

Examples for the creep of Meissner holes can be seen in
\figurenames~\ref{fig:fluxcreep2}((a)--(f)). There, remagnetization of
superconducting plates in normal fields occurs by closed induction
loops centered at the remagnetization front~\cite{VWCGKN97}. This
formation of the Meissner holes is a consequence of changing the field
distribution at the edge of a plate in the normal field. The
measurements of the fluxons in the graphite sample are shown in
\figurenames~\ref{fig:fluxcreep2}((g)--(h)), where time-dependent,
thermally activated creep gives rise to a weak dissipation.

\section{Summary and Conclusion}
\label{sec:conclusion}

We have prepared and measured a natural graphite flake (Sri Lanka),
which shows a transition at $T\simeq370$~K in the electrical resistance
and its remanence. We were able to localize a persistent current using
MFM, which is stable for several weeks, yet shows signs of
creep. Because of the apparent similarity of the line phase scans to
the scans one would obtain around magnetic domain walls, we summarize
below the reasons that speak against this possible origin.

\paragraph{Reasons against a magnetic origin} The first experimental
observation at odds with a magnetic origin, is that the phase is
constant over all sample when it is measured in the virgin state, that
means after cooling down from above $T_c$ without applied magnetic
field. Would there be a magnetic order transition, it is highly
unlikely that there were no measurable stray fields from magnetic
domains, as \textit{all} domains would have to be aligned in the same
direction. Even if these domains were smaller than the spatial
resolution, some unresolved signal should have been measured.

We have measured the phase feature at the apparent current path over a
large area with a length of $\approx600~\upmu$m. It is very unlikely
to obtain such a large domain wall in remanence and after application
of a field as low as $30$~mT at room temperature. If that was the
case, the exchange interaction would be strong enough to form large
magnetic domains when cooling below the Curie temperature at no
applied field, which is clearly not the case.

The sudden vanishing of the phase, i.e.~the stray fields, at
$T\approx370$~K is not what one would expect when approaching the
Curie temperature, where a continuous decreasing amount of magnetically
ordered moments  results in a decreasing phase shift and/or
more pronounced domain wall structure.

As discussed in~\cite{PECQZL16}, there is no evidence for
magnetic order in all measured samples. Neither the amount of
impurities is significant, nor the intrinsic disorder, in very
different samples, ranging from bulk to multilayered graphene flakes.

The measured scan profile of the phase at the path border is not
compatible with the edge of a magnetically ordered domain and/or a
domain wall. Part of the scan profile, i.e.~the zero crossing and the
sign change between the left and right side of the border, appears to
be compatible with a N{\'e}el wall between two magnetic domains. But,
the measured line scan indicates that a homogeneous field remains only
in the interior of the loop, which contradicts the expected profile
from the two antiparallel domains (after application of an external
field perpendicular to the sample surface). Further, it is unlikely
that it is a single domain, because parts of the same material would
couple ferromagnetically while other areas do not.  In addition, this
asymmetry of the phase border implies that even if the domains were
pointing all in the same direction (the virgin state) there would have
to be stray fields at this position.  This was not the case in the
virgin measurements and after zero field cool. It is also highly
unlikely that those magnetic domains would be produced only after
applying a field normal to the graphene planes and that all are
antiparallel aligned in such a long path.

The observed change in the position of the current path with time
rules out a correlation with any magnetic topographic feature, such as
zigzag edges or conglomeration of magnetic impurities. This change was
not only observed in the measurements presented above, but was
recognizable during all measurement time. This was especially imminent
when we returned to the starting position, after we reached the
rough edge region, where we were not able to continue the MFM
measurements, but had to continue the experiments on the other side of
the found current path.

The virgin state is not reached anymore, even changing the amplitude
direction of the maximum applied field, see~\cite{PECQZL16}. Further,
the magnetoresistance is positive and large at fields of a few mT, in
contrast to the negative and weak magnetoresistance at $T\geq300$~K for
magnetic graphite~\cite{chap3}.

Our results indicate the presence of a trapped flux through a
persistent current, which we interpret as superconductivity. Further,
MFM, as well as other scanning magnetic imaging techniques, can be
used to identify the regions of interest of the graphite sample. This
will undoubtedly assist to further characterize the
superconducting-like interfaces in graphite, paving the way for
their future device implementations.

\begin{acknowledgements}
  We thank Henning Beth for providing us with the natural graphite
  sample from Sri Lanka and A. Setzer for the XRD
  characterization. C.E.P. gratefully acknowledges the support
  provided by The Brazilian National Council for the Improvement of
  Higher Education (CAPES). M.S. and J.B.Q. are supported by the DFG
  collaboration project SFB762.
\end{acknowledgements}
\bibliography{bibliography}   

\end{document}